\documentclass[useAMS,usenatbib]{mn2e}
\usepackage{txfonts}
\usepackage{graphicx} 
\usepackage{rotating}
\usepackage{caption}

\title[How to take the interstellar weather with you]{How to take the interstellar weather with you in pulsar timing analysis}
\author[L. Lentati et al.]{\parbox{\textwidth}{L. Lentati$^{1}$\thanks{E-mail:
ltl21@cam.ac.uk}, P. Alexander$^{1}$, M. P.  Hobson$^{1}$\vspace{0.4cm}}\\ %
$^{1}$Astrophysics Group, Cavendish Laboratory, JJ Thomson Avenue,  Cambridge, CB3 0HE, UK}

\begin{document}

\maketitle

\label{firstpage}

\begin{abstract}
Here we present a Bayesian method of including discrete measurements of dispersion measure due to the interstellar medium in the direction of a pulsar as prior information in the analysis of that pulsar. We use a simple simulation to show the efficacy of this method, where the inclusion of the additional measurements results in both a significant increase in the precision with which the timing model parameters can be obtained, and an improved upper limit on the amplitude of any red noise in the dataset.  We show that this method can be applied where no multi-frequency data exists across much of the dataset, and where there is no simultaneous multi-frequency data for any given observing epoch.  Including such information in the analysis of upcoming International Pulsar Timing Array (IPTA) and European Pulsar Timing Array (EPTA) data releases could therefore prove invaluable in obtaining the most constraining limits on gravitational wave signals within those datasets.

\end{abstract}

\begin{keywords}
methods: data analysis, pulsars: general, pulsars:individual
\end{keywords}

\section{Introduction}

The first direct detection of gravitational waves is a key science mission around the world, with many different approaches being advocated.  These include ground and space-based laser interferometers \citep{2010CQGra..27h4006H, 2012CQGra..29l4016A}, and pulsar timing arrays (collections of galactic millisecond pulsars (MSPs); \citealt{1990ApJ...361..300F}), and it is with the latter of these that we will be concerned with here.

It is the exceptional stability of MSPs, with decade long observations providing timing measurements that show fractional instabilities similar to atomic clocks (e.g. \citealt{1994ApJ...428..713K,1997A&A...326..924M}), that makes them key to the pursuit of a wide range of scientific endeavors. For example, observations of the pulsar PSR B1913+16 provided the first indirect detection of gravitational waves \citep{1989ApJ...345..434T}, whilst the double pulsar system PSR J0737-3039A/B provides precise measurements of several `post Keplerian' parameters allowing for additional stringent tests of general relativity \citep{2006Sci...314...97K}.

Current theoretical limits (\citealt{2013MNRAS.433L...1S}) place the amplitude of a stochastic gravitational wave background (GWB) generated by coalescing black holes (e.g. \citealt{2003ApJ...583..616J,2001astro.ph..8028P}) at only a factor 3-10 lower than current observational limits (e.g.  \citealt{2011MNRAS.414.3117V}).  In order to make the first tentative detections of these signals as much ancillary data will be needed as possible in order to constrain the other components present in the data.

Dispersion measure (DM) variations are thought to be one of the largest components of noise in pulsar timing data (e.g. \citealt{2011PhRvD..83h1301J}), and many different methods exist to describe it (e.g. \citealt{2013arXiv1310.2120L}, henceforth L13; Lee et al. submitted 2013; \citealt{2013ApJ...762...94D,2013MNRAS.429.2161K}).  In the near future, observations from LOFAR \citep{2013A&A...556A...2V} will allow precise measurements of DM in the direction of pulsars to be used in PTA analysis.  Including this information in subsequent analysis in order to constrain the DM signal and separate it from the gravitational waves will thus be critical. 

In this article we describe how to include such DM measurements as prior information in pulsar timing analysis in order to constrain the signal realisation for the DM by modifying the existing Bayesian techniques presented in L13.  In section \ref{Section:Bayes} we give a brief overview of our Bayesian methodology, and in Section \ref{Section:Models} derive the likelihood that we use to include the additional DM measurements when analysing the simulated data in Section \ref{Section:Sims}.  Finally we will provide some concluding remarks in Section \ref{Section:Conclusion}.

This research is the result of the common effort to directly detect gravitational waves using pulsar timing, known as the European Pulsar Timing Array (EPTA) \citet{2008AIPC..983..633J} \footnote{www.epta.eu.org/}.

\section{Bayesian Inference}
\label{Section:Bayes}

Our method for performing pulsar timing analysis is built upon the principles of Bayesian inference, which provides a consistent approach to the estimation of a set of parameters $\Theta$ in a model or hypothesis $H$ given the data, $D$.  Bayes' theorem states that:

\begin{equation}
\mathrm{Pr}(\Theta \mid D, H) = \frac{\mathrm{Pr}(D\mid \Theta, H)\mathrm{Pr}(\Theta \mid H)}{\mathrm{Pr}(D \mid H)},
\end{equation}
where $\mathrm{Pr}(\Theta \mid D, H) \equiv \mathrm{Pr}(\Theta)$ is the posterior probability distribution of the parameters,  $\mathrm{Pr}(D\mid \Theta, H) \equiv L(\Theta)$ is the likelihood, $\mathrm{Pr}(\Theta \mid H) \equiv \pi(\Theta)$ is the prior probability distribution, and $\mathrm{Pr}(D \mid H) \equiv Z$ is the Bayesian Evidence.

In parameter estimation, the normalizing evidence factor is usually ignored, since it is independent of the parameters $\Theta$.   Inferences are therefore obtained by taking samples from the (unnormalised) posterior using, for example, standard Markov chain Monte Carlo (MCMC) sampling methods.  

An alternative to MCMC is the nested sampling approach \citep{2004AIPC..735..395S}, a Monte-Carlo method targeted at the efficient calculation of the evidence, that also produces posterior inferences as a by-product.  In \cite{2009MNRAS.398.1601F} and \cite{2008MNRAS.384..449F} this nested sampling framework was built upon with the introduction of the MultiNest algorithm, which provides an efficient means of sampling from posteriors that may contain multiple modes and/or large (curving) degeneracies, and also calculates the evidence.  Since its release MultiNest has been used successfully in a wide range of astrophysical problems, from detecting the Sunyaev-Zel'dovich effect in galaxy clusters \citep{2012arXiv1210.7771C}, to inferring the properties of a potential stochastic gravitational wave background in pulsar timing array data \citep{2013PhRvD..87j4021L}.  In the following sections we make use of the MultiNest algorithm to obtain our estimates of the posterior probability distributions for both timing model, and stochastic parameters.

Recently TempoNest (L13) was introduced as a means of performing a simultaneous analysis of either the linear or non-linear timing model and additional stochastic parameters using MultiNest to efficiently explore this joint parameter space, whilst using TEMPO2 \citep{2006MNRAS.369..655H, 2006MNRAS.372.1549E, 2009MNRAS.394.1945H} as an established means of evaluating the timing model at each point in that space.  We incorporate the likelihood developed in section \ref{Section:Models} into TempoNest in order to perform the analysis described in Section \ref{Section:Sims}.

\section{Pulsar timing likelihood}
\label{Section:Models}

\subsection{Timing model and white noise parameters}
For any pulsar we can write the TOAs for the pulses as a sum of both a deterministic and a stochastic component:

\begin{equation}
\bmath{t}_{\mathrm{tot}} = \bmath{t}_{\mathrm{det}} + \bmath{t}_{\mathrm{sto}},
\end{equation}
where $\bmath{t}_{\mathrm{tot}}$ represents the $n$ TOAs for a single pulsar, with $\bmath{t}_{\mathrm{det}}$ and $\bmath{t}_{\mathrm{sto}}$ the deterministic and stochastic contributions to the total respectively, where any contributions to the latter will be modelled as random Gaussian processes.  
Writing the deterministic signal due to the timing model as $\bmath{\tau}(\bmath{\epsilon})$, and the uncertainty associated with a particular TOA as:

\begin{equation}
\hat{\sigma}_i^2 = (\alpha_i\sigma_i)^2 + \beta_i^2,
\end{equation}
where $\alpha$ and $\beta$ represent the EFAC and EQUAD parameters applied to TOA $i$ respectively, we can write the probability that the data is described by the timing model parameters $\bmath{\epsilon}$ and white noise parameters $\alpha$ and $\beta$ as:

\begin{equation}
\label{Eq:WhiteLike}
\mathrm{Pr}(\bmath{t} | \bmath{\epsilon}, \bmath{\alpha}, \beta) \propto \left(\prod_{i=1}^n\hat{\sigma}_i^2\right)^{-\frac{1}{2}}\exp{\left(-\frac{1}{2}\sum_{i=1}^n\frac{(t_i - \tau(\bmath{\epsilon})_i)^2}{\hat{\sigma}_i^2}\right)}.
\end{equation}

\subsection{Dispersion measure variations}

To include the DM variations, which we will denote $\bmath{t}_{\mathrm{DM}}$, we begin by following the same process as in L13. Writing it in terms of its Fourier coefficients $\bmath{a}$ so that $\bmath{t}_{\mathrm{DM}} = \bmath{F}\bmath{a}$ where $\bmath{F}$ denotes the Fourier transform such that for frequency $\nu_s$ and time $t$ we will have both:

\begin{equation}
\label{Eq:FMatrix}
F(\nu_s,t) = \frac{1}{T\kappa\nu_o(t)^2}\sin\left(2\pi\nu_s t\right),
\end{equation}
and an equivalent cosine term.  Here the dispersion constant $\kappa$ is given by:

\begin{equation}
\kappa \equiv 2.41 \times 10^{-16}~\mathrm{Hz^{-2}~cm^{-3}~pc~s^{-1}},
\end{equation}
$T$ is the total observing timespan, $\nu_o(t)$ is the observing frequency for the TOA at barycentric arrival time $t$, and $\nu_s$ the frequency of the signal to be sampled.  Defining the number of coefficients to be sampled by $n_{\mathrm{max}}$, we can then include the set of frequencies with values $\nu_s=n/T$, where $n$ extends from 1 to $n_{\mathrm{max}}$.
For typical PTA data \cite{2012MNRAS.423.2642L} show that for frequency independent spin noise, a low frequency cut off of $1/T$ is sufficient to accurately describe the expected long term variations present in the data, as the quadratic included in the timing model in the form of the spindown parameters acts as a proxy to lower frequency signals. 
For DM variations, however, these terms must be accounted for either by explicitly including these low frequencies in the model, or by including a quadratic in DM to act as a proxy, as with the red noise, defined as: 

\begin{equation}
 Q_{\mathrm{DM}}(t_i)= \Delta_0D(t_i) + \Delta_1 t_iD(t_i) + \Delta_2 t_i^2D(t_i)
\end{equation}
with $\Delta_{0,1, 2}$ free parameters to be fitted, $t_i$ the barycentric arrival time for TOA $i$ and $D(t_i)$ elements of a vector:

\begin{equation}
D(t_i) = 1/(\kappa\nu^2_o(t_i)).
\end{equation}  

For a single pulsar the covariance matrix $\bmath{\varphi}$ of the Fourier coefficients $\bmath{a}$ will be diagonal, with components

\begin{equation}
\label{Eq:BPrior}
\varphi_{ij} = \left< a_ia_j^*\right> = \varphi_{i}\delta_{ij},
\end{equation}
where there is no sum over $i$, and the set of coefficients $\{\varphi_{i}\}$ represent the theoretical power spectrum for the DM variations in the residuals.  

As discussed in \cite{2013PhRvD..87j4021L}, whilst Eq \ref{Eq:BPrior} states that the Fourier modes are orthogonal to one another, this does not mean that we assume they are orthogonal in the time domain where they are sampled, and it can be shown that this non-orthogonality is accounted for within the likelihood.  Instead, in Bayesian terms, Eq. \ref{Eq:BPrior} represents our prior knowledge of the power spectrum coefficients within the data.  We are therefore stating that, whilst we do not know the form the power spectrum will take, we know that the underlying Fourier modes are still orthogonal by definition, regardless of how they are sampled in the time domain.  It is here then that, should one wish to fit a specific model to the power spectrum coefficients at the point of sampling, such as a broken, or single power law, the set of coefficients $\{\varphi_{i}\}$ should be given by some function $f(\Theta)$, where we sample from the parameters $\Theta$ from which the power spectrum coefficients $\{\varphi_{i}\}$ can then be derived.

We can then write the joint probability density of the timing model, white noise parameters, power spectrum coefficients and the signal realisation, Pr$(\bmath{\epsilon}, \bmath{\alpha}, \beta, \{\varphi_i\}, \bmath{a} \;|\; \bmath{t})$, as:

\begin{eqnarray}
\label{Eq:Prob}
\mathrm{Pr}(\bmath{\epsilon}, \bmath{\alpha}, \beta, \{\varphi_i\}, \bmath{a} \;|\; \bmath{t}) \; &\propto& \; \mathrm{Pr}(\bmath{t} |  \bmath{\epsilon}, \bmath{\alpha}, \beta, \bmath{a}) \; \\\ \nonumber
&\times & \mathrm{Pr}(\bmath{a} | \{\varphi_i\}, \mathrm{\pi_{DM}}) \; \mathrm{Pr}(\{\varphi_i\}), \nonumber
\end{eqnarray}
where $\pi_{DM}$ represents any additional prior information regarding the DM signal realisation.  Henceforth we will consider $\pi_{DM}$ to be given by a vector of measurements of the DM at some set of arbitrary times with associated measurement errors, which we will denote $\bmath{L_{DM}}$ and $\bmath{\sigma_{DM}}$ respectively. For our choice of $\mathrm{Pr}(\{\varphi_i\})$ we use an uninformative prior that is uniform in $\log_{10}$ space, and draw our samples from the parameter $\rho_i = \log_{10}(\varphi_i)$ instead of $\varphi_i$.  Given this choice of prior the conditional distributions that make up Eq. \ref{Eq:Prob} can be written:

\begin{eqnarray}
\label{Eq:ProbTime}
& &\mathrm{Pr}(\bmath{t} |\bmath{\epsilon}, \bmath{\alpha}, \beta, \bmath{a}) \; \propto \; \frac{1}{\sqrt{\mathrm{det}(\bmath{N})}}  \\
& \times & \exp\left[-\frac{1}{2}(\bmath{t} - \bmath{f_{DM}} - \bmath{\tau}(\bmath{\epsilon}) - \bmath{F}\bmath{a})^T\bmath{N}^{-1}(\bmath{t} -  \bmath{f_{DM}} - \bmath{\tau}(\bmath{\epsilon}) - \bmath{F}\bmath{a})\right] \nonumber
\end{eqnarray}
where $\bmath{N}_{ij} = \hat{\sigma}^2_{i}\delta_{ij}$ and represents the white noise errors in the residuals, and $\bmath{f_{DM}}$ describes components of the DM model in addition to those contained in the Fourier modes, such as the quadratic terms in the timing model which we have separated from the $\bmath{\tau}(\bmath{\epsilon})$ term for clarity.  We then also have:

\begin{eqnarray}
\label{Eq:ProbFreq}
\mathrm{Pr}(\bmath{a} | \{\rho_i\}, \mathrm{\bmath{L_{DM}}}) \; &\propto& \; \frac{1}{\sqrt{\mathrm{det}\bmath{\varphi}}} \exp\left[-\frac{1}{2}\bmath{a}^{*T}\bmath{\varphi}^{-1}\bmath{a}\right] \\ \nonumber
&\times& \frac{1}{\sqrt{\mathrm{det}\bmath{\psi}}}\exp\left[-\frac{1}{2}(\bmath{L_{DM}}-\bmath{f_{DM}}-\bmath{F_La})^T\bmath{\psi}^{-1}\right.\nonumber \\
&\times& \left.(\bmath{L_{DM}}-\bmath{f_{DM}}-\bmath{F_La})\right].
\end{eqnarray}
Here $\bmath{F_L}$ is a matrix of Fourier modes as in Eq. \ref{Eq:FMatrix}, however with points evaluated at the times that additional DM measurements were made, and $\bmath{\psi}$ is the diagonal noise matrix for the additional DM measurements with values $\bmath{\sigma_{DM}}$.  We then marginalise over all Fourier coefficients $\bmath{a}$ analytically in order to find the posterior for the remaining parameters alone.

In order to perform the marginalisation over the Fourier coefficients $\bmath{a}$, we first write the log of the likelihood in Eq \ref{Eq:Prob}, which, excluding the determinant terms, and denoting $(\bmath{t} - \bmath{f_{DM}} - \bmath{\tau}(\bmath{\epsilon}))$ as $\bmath{\delta t}$, $(\bmath{L_{DM}}-\bmath{f_{DM}})$ as $\bmath{\delta_{DM}}$, $(\bmath{F}^T\bmath{N}^{-1}\bmath{F} + \bmath{F_L}^T\bmath{\psi}^{-1}\bmath{F_L} + \bmath{\varphi}^{-1})$ as $\bmath{\Sigma}$ and $(\bmath{F}^T\bmath{N}^{-1}\bmath{\delta t} + \bmath{F_L}^T\bmath{\psi}^{-1}\bmath{\delta_{DM}}) $ as $\bmath{d}$ is given by:
\begin{equation}
\label{Eq:LogL}
\log \mathrm{L} = -\frac{1}{2} \bmath{\delta t}^T\bmath{N}^{-1} \bmath{\delta t} - \frac{1}{2} \bmath{\delta_{DM}}^T\bmath{\psi}^{-1}\bmath{\delta_{DM}} - \frac{1}{2}\bmath{a}^T\bmath{\Sigma}\bmath{a} + \bmath{d}^T\bmath{a}.
\end{equation}
Taking the derivitive of $\log \mathrm{L}$ with respect to $\bmath{a}$ gives us:
\begin{equation}
\label{Eq:Grada}
\frac{\partial \log \mathrm{L}}{\partial \bmath{a}} =  -\bmath{\Sigma}\bmath{a} + \bmath{d},
\end{equation}
which can be solved to give us the maximum likelihood vector of coefficients $\hat{\bmath{a}}$:
\begin{equation}
\label{Eq:amax}
\hat{\bmath{a}} = \bmath{\Sigma}^{-1}\bmath{d}.
\end{equation}
Re-expressing Eq. \ref{Eq:LogL} in terms of $\hat{\bmath{a}}$:

\begin{eqnarray}
\log \mathrm{L} &=& -\frac{1}{2} \bmath{\delta t}^T\bmath{N}^{-1} \bmath{\delta t} - \frac{1}{2} \bmath{\delta_{DM}}^T\bmath{\psi}^{-1}\bmath{\delta_{DM}}  \nonumber \\
& + &  \frac{1}{2}\hat{\bmath{a}}^T\bmath{\Sigma}\hat{\bmath{a}} - \frac{1}{2}(\bmath{a} - \hat{\bmath{a}})^T\bmath{\Sigma}(\bmath{a} - \hat{\bmath{a}}),
\end{eqnarray}
the 3rd term in this expression can then be integrated with respect to the $m$ elements in $\bmath{a}$ to give:
\begin{eqnarray}
I &=& \int_{-\infty}^{+\infty}\mathrm{d}\bmath{a}\exp\left[-\frac{1}{2}(\bmath{a} - \hat{\bmath{a}})^T\bmath{\Sigma}(\bmath{a} - \hat{\bmath{a}})\right] \nonumber \\
&=& (2\pi)^m~\mathrm{det} ~ \bmath{\Sigma}^{-\frac{1}{2}}.
\end{eqnarray}
Our marginalised probability distribution for a set of GWB coefficients is then given as:
\begin{eqnarray}
\label{Eq:Margin}
\mathrm{Pr}(\{\varphi_i\} \;|\; \bmath{\delta t}) &\propto& \frac{\mathrm{det} \left(\bmath{\Sigma}\right)^{-\frac{1}{2}}}{\sqrt{\mathrm{det} \left(\bmath{\varphi}\right)~\mathrm{det}\left(\bmath{N}\right)~\mathrm{det}\left(\mathbf{\psi}\right)}} \\
&\times&\exp\left[-\frac{1}{2}\left(\bmath{\delta t}^T\bmath{N}^{-1} \bmath{\delta t} + \frac{1}{2} \bmath{\delta_{DM}}^T\bmath{\psi}^{-1}\bmath{\delta_{DM}} - \bmath{d}^T\bmath{\Sigma}^{-1}\bmath{d}\right)\right], \nonumber
\end{eqnarray}
\subsection{Changing the weight of the additional DM measurements}

In much the same way in which we include an EFAC in the white noise matrix to alter the weighting of the data, we can include an EFAC in the matrix  $\bmath{\psi}$ in order to alter the weight of the DM measurements in the event that the errors provided are under, or over estimated, or if the measurements are not consistent with the timing data.  As such we can simply define:

\begin{equation}
\bmath{\sigma'_{DM}} = \alpha_{DM}\bmath{\sigma_{DM}}
\end{equation}
In the simulations described in Section \ref{Section:Sims} however, we will not include this additional parameter.

\subsection{Including additional red spin noise}

We can include additional, frequency independent red `spin' noise in much the same way as the DM variations.  As before we define a matrix of Fourier modes for a set of $n$ frequencies:

\begin{equation}
\label{Eq:FRMatrix}
F_R(\nu_s,t) = \frac{1}{T}\sin\left(2\pi\nu_s t\right),
\end{equation}
and an equivalent cosine term.  These rows can then be appended to the Fourier matrix in Eq.\ref{Eq:FMatrix}, which we will denote here $F_{DM}$ to form a new matrix containing both the red noise and DM terms:

\begin{equation}
F = {F_{DM} \choose F_{R}}.
\end{equation} 
Similarly the matrix $\bmath{\varphi}$ is extended to accommodate the new power spectrum coefficients required to describe the spin noise.  The additional DM prior term is then kept the same: in forming the matrix $\Sigma$ we add the term $\bmath{F_L}^T\bmath{\psi}^{-1}\bmath{F_L}$ to only the section of the matrix $(\bmath{F}^T\bmath{N}^{-1}\bmath{F} + \bmath{\varphi}^{-1})$ that corresponds to the autocorrelated terms of the DM modes, and similarly the vector $\bmath{F_L}^T\bmath{\psi}^{-1}\bmath{\delta_{DM}}$ is added only to the part of $\bmath{F}^T\bmath{N}^{-1}\bmath{\delta t}$ concerned with the DM Fourier modes when forming $\bmath{d}$.

\subsection{Analytical marginalisation over the timing model}
\label{Section:Margin}

In many pulsar timing datasets, phase jumps are fitted between different groups of observations, or there might be other parameters that are not of interest, such as a constant phase offset, resulting in a potentially significant increase in the number of parameters to be fit for in analysis. If the specific values of such parameters are not of importance we can marginalise analytically over them, greatly reducing the dimensionality of the problem.

If we separate the timing model into a contribution from the set of parameters that we wish to parameterise $\bmath{\tau}(\bmath{\epsilon})$ and a contribution from the set of $m$ parameters that we plan to marginalise over analytically $\bmath{\tau}(\bmath{\epsilon'})$  then we can write the probability that the data $\bmath{t}$ is described by the remaining parameters $\bmath{\epsilon}$ and any additional parameters $\bmath{\theta}$ we wish to include as:

\begin{eqnarray}
\mathrm{Pr}(\bmath{t} | \bmath{\epsilon}, \bmath{\theta}) &=& \int \; \mathrm{d}^m\bmath{\epsilon'} \; \mathrm{Pr}(\bmath{\epsilon'}) \;\mathrm{Pr}(\bmath{t} | \bmath{\epsilon'},\bmath{\epsilon},\bmath{\theta}).
\end{eqnarray}
Using a uniform prior on the $m$ $\bmath{\epsilon'}$  parameters, we use the same approach as described in  \citep{2013MNRAS.428.1147V} to perform this marginalisation process analytically.  This results in equation \ref{Eq:LLikeMargin}, which is the expression we will be using in the subsequent analysis:

\begin{eqnarray}
\label{Eq:LLikeMargin}
\mathrm{Pr}(\bmath{\epsilon}, \bmath{\alpha}, \beta, \{\varphi_i\} | \bmath{t}) &\propto& \frac{\mathrm{det} \left(\bmath{\Sigma}\right)^{-\frac{1}{2}}}{\sqrt{\mathrm{det} \left(\bmath{\varphi}\right)~\mathrm{det}\left(\bmath{\hat{N}}\right)~\mathrm{det}\left(\mathbf{\psi}\right)}} \\
&\times&\exp\left[-\frac{1}{2}\left(\bmath{\delta t}^T\bmath{\bmath{\hat{N}}}^{-1} \bmath{\delta t} + \frac{1}{2} \bmath{\delta_{DM}}^T\bmath{\psi}^{-1}\bmath{\delta_{DM}} \right.\right. \nonumber \\
&-& \left.\left.\bmath{\hat{d}}^T\bmath{\hat{\Sigma}}^{-1}\bmath{\hat{d}}\right)\right], 
\end{eqnarray}
where $\bmath{\hat{N}} = \bmath{G}(\bmath{G}^T\bmath{NG})^{-1}\bmath{G}^T$, $\bmath{\hat{\Sigma}} = (\bmath{F}^T{\bmath{\hat{N}}}^{-1}\bmath{F} + \bmath{F_L}^T\bmath{\psi}^{-1}\bmath{F_L} + \bmath{\varphi}^{-1})$ and $\bmath{\hat{d}} = \bmath{F}^T\bmath{\hat{N}}^{-1}\bmath{\delta t}+\bmath{F_L}^T\bmath{\psi}^{-1}\bmath{\delta_{DM}}$.

\section{Application to simulated data}
\label{Section:Sims}

\begin{table*}
\caption{Parameter estimates for PSR J0030+0451 with and without additional prior information on DM.  $^u$ indicates a 2$\sigma$ upper limit.}
\begin{tabular}{lccc}
\hline
Model Parameter & Injected value & with DM prior & without DM prior \\
\hline
Right ascension, $\alpha$\dotfill & 00:30:27.4302  &  00:30:27.43036(13) & 00:30:27.4302(3)\\ 
Declination, $\delta$\dotfill & +04:51:39.7402 & +04:51:39.738(4) & +04:51:39.740(12)\\ 
Pulse frequency, $\nu$ (s$^{-1}$)\dotfill  & 205.53069608813808 & 205.530696088139(4) & 205.530696088153(3)\\ 
First derivative of pulse frequency, $\dot{\nu}$ (s$^{-2}$)\dotfill & $-$4.30100$\times 10^{-16}$  &  $-$4.3011(4)$\times 10^{-16}$ & $-$4.303(2)$\times 10^{-16}$\\ 
DM (cm$^{-3}$ pc)\dotfill					& 0.01 & 0.010(2) & 0.007(4)\\
First derivative of dispersion measure, $\dot{DM}$ (cm$^{-3}$pc\,yr$^{-1}$)\dotfill & - & $-$0.002(2)& 0.001(3)\\
Second derivative of dispersion measure, $\ddot{DM}$ (cm$^{-3}$pc\,yr$^{-2}$)\dotfill & - & 0.0006(5)& $-$0.0001(5)\\	
Proper motion in right ascension, $\mu_{\alpha}$ (mas\,yr$^{-1}$)\dotfill & $-$5.3932 & $-$5.8(6) & $-$5.7(15)\\ 
Proper motion in declination, $\mu_{\delta}$ (mas\,yr$^{-1}$)\dotfill & $-$1.7401 & $-$0.7(14) & $-$1(3)\\ 
Parallax, $\pi$ (mas)\dotfill & 4.207 & 4.44(15) & 4.8(3)\\ 
$\log_{10} \mathrm{A_{DM}}$ \dotfill & 0.699 &0.70(6) &0.68(6) \\
$\gamma_{\mathrm{DM}}$ \dotfill & 1.7 & 1.50(13) &1.49(17)   \\
$\mathrm{A_{red}}$ \dotfill & - & 0.0011$^u$ & 0.003$^u$  \\
$\gamma_{\mathrm{red}}$ \dotfill & - & 2.2(13) & 2.4(15)\\

\hline
\hline
\end{tabular}
\label{Table:Sim1}
\end{table*}

In order to check the efficacy of the method described in Section \ref{Section:Models} we simulate a dataset for the isolated pulsar J0030+0451.  We begin by taking the injected parameter values given in Table \ref{Table:Sim1} and generate a 5 year dataset with uneven sampling in the time domain, but with an average cadence of $\sim$ 2 weeks.  We include two observing frequencies at 1440 and 2440 MHz where the higher frequency observations exist only for the latter $\sim 2/3$ of the observations, and where no multi-frequency data exists for any given observing epoch of duration $\sim$ 2 weeks.  We then add variations in the DM that are described by a power law with functional form $S(\nu) = A_{DM}^2\nu^{-\gamma_{DM}}$ with a spectral index of $\gamma_{DM}=1.7$.  Note that we do not list injected values for the $DM1$ and $DM2$ parameters as these are used simply as proxies to the low frequency DM variations, and as such we do not know a priori what these values will take.

\begin{figure*}
\begin{center}$
\begin{array}{cc}
\includegraphics[width=80mm]{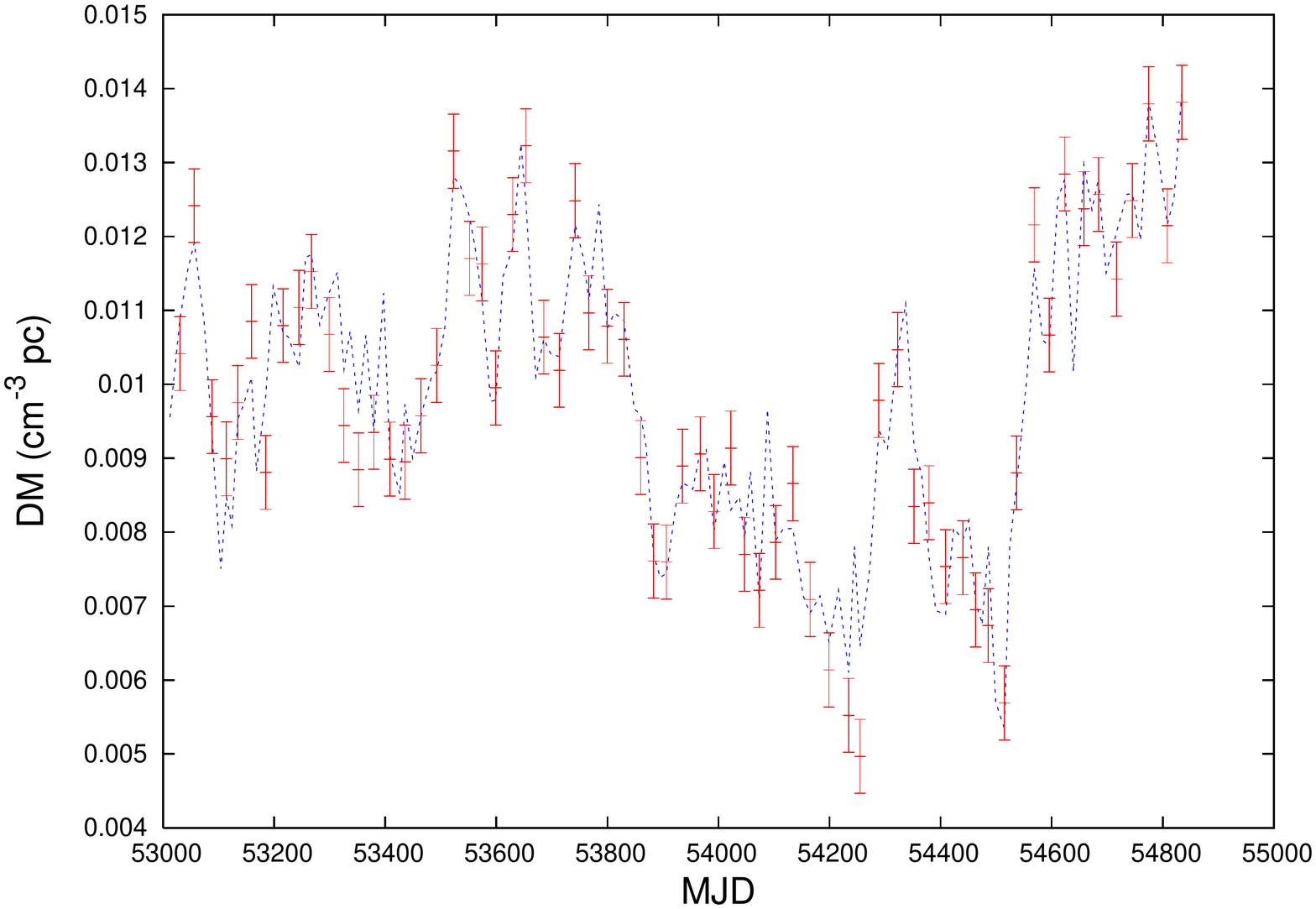} &
\includegraphics[width=80mm]{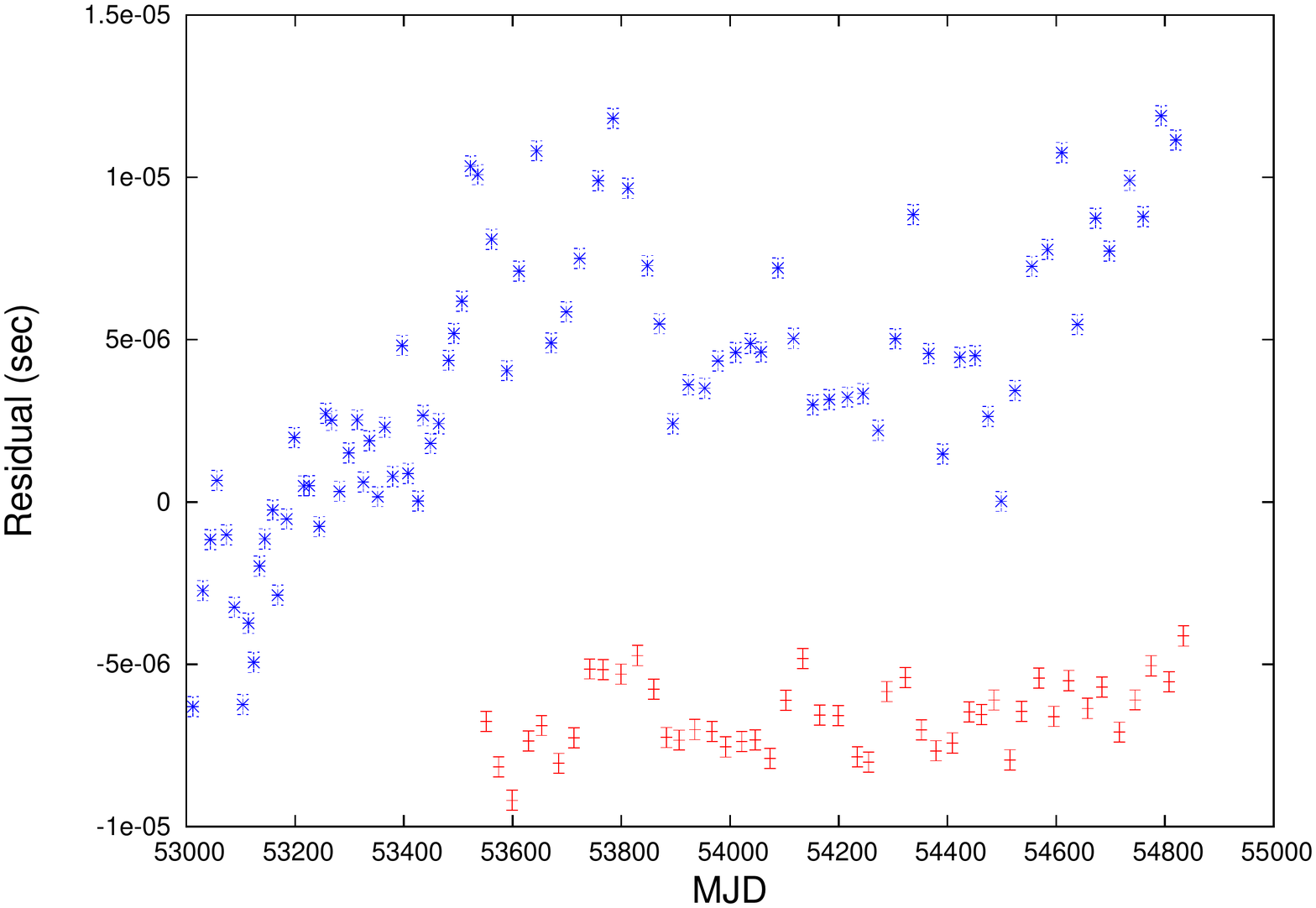} \\
\end{array}$
\end{center}
\vspace{-1cm}
\caption{(left) The injected DM signal used in the simulation for PSR J0030+0451 (blue line) and the simulated DM observations sampled monthly (red points).  (right) The timing residuals after subtracting the timing model given in Table \ref{Table:Sim1}.  The colours indicate the two observing frequencies included in the simulation of 1440 (blue) and 2440 (red) MHz.\label{Fig:Res}}
\end{figure*}

We then simulate discrete observations of the total DM signal that will act as our prior, $\bmath{L_{DM}}$.  We generate monthly samples that are scattered around the true signal with an rms of $\sigma_{DM} = 0.005$cm$^{-3}$~pc. The injected DM signal, simulated DM observations, and the final residuals obtained when subtracting the injected timing model parameters in Table \ref{Table:Sim1} except the DM parameters, are shown in Fig. \ref{Fig:Res}.

\begin{figure*}
\begin{center}$
\begin{array}{cc}
\includegraphics[width=80mm]{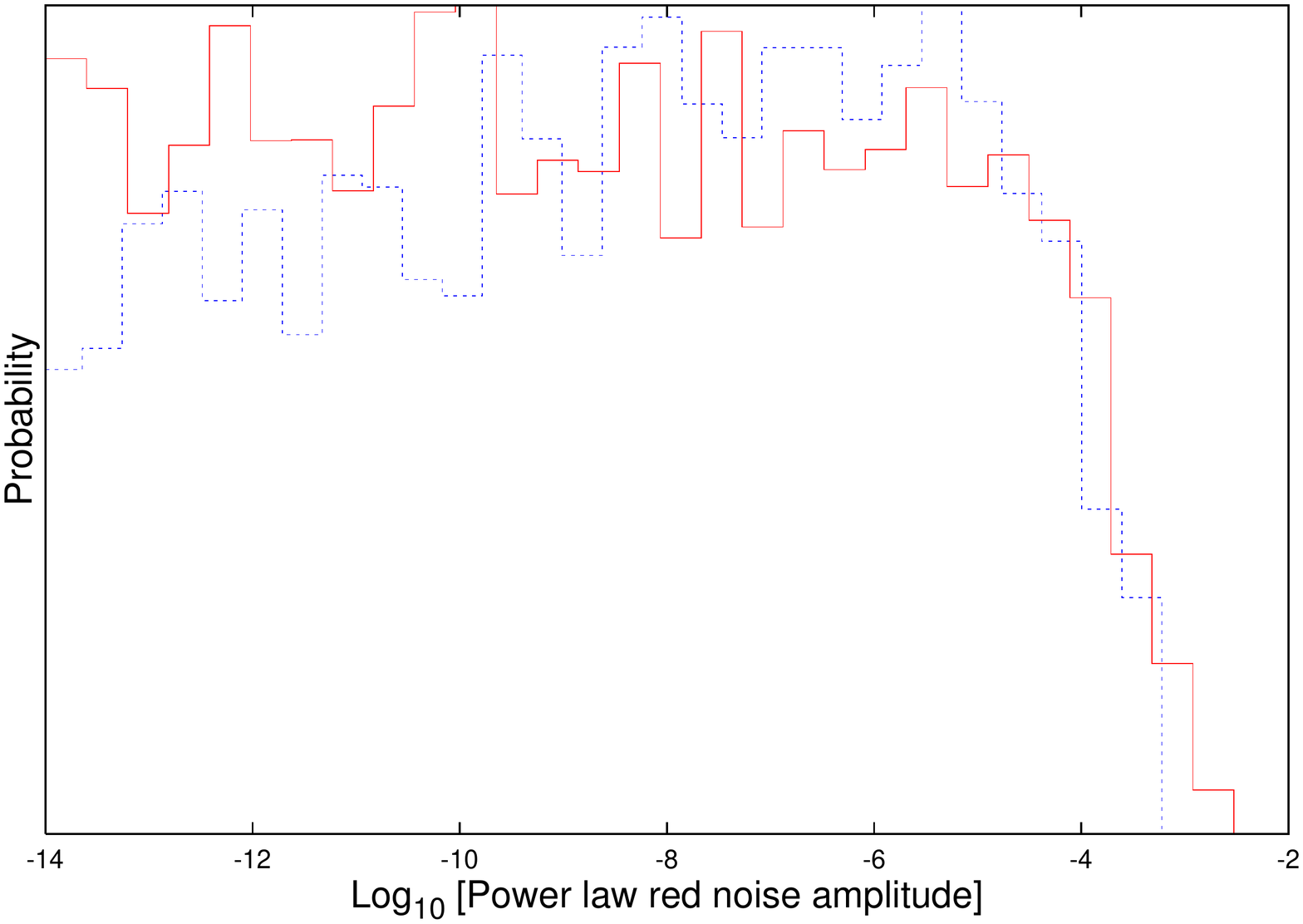} &
\includegraphics[width=80mm]{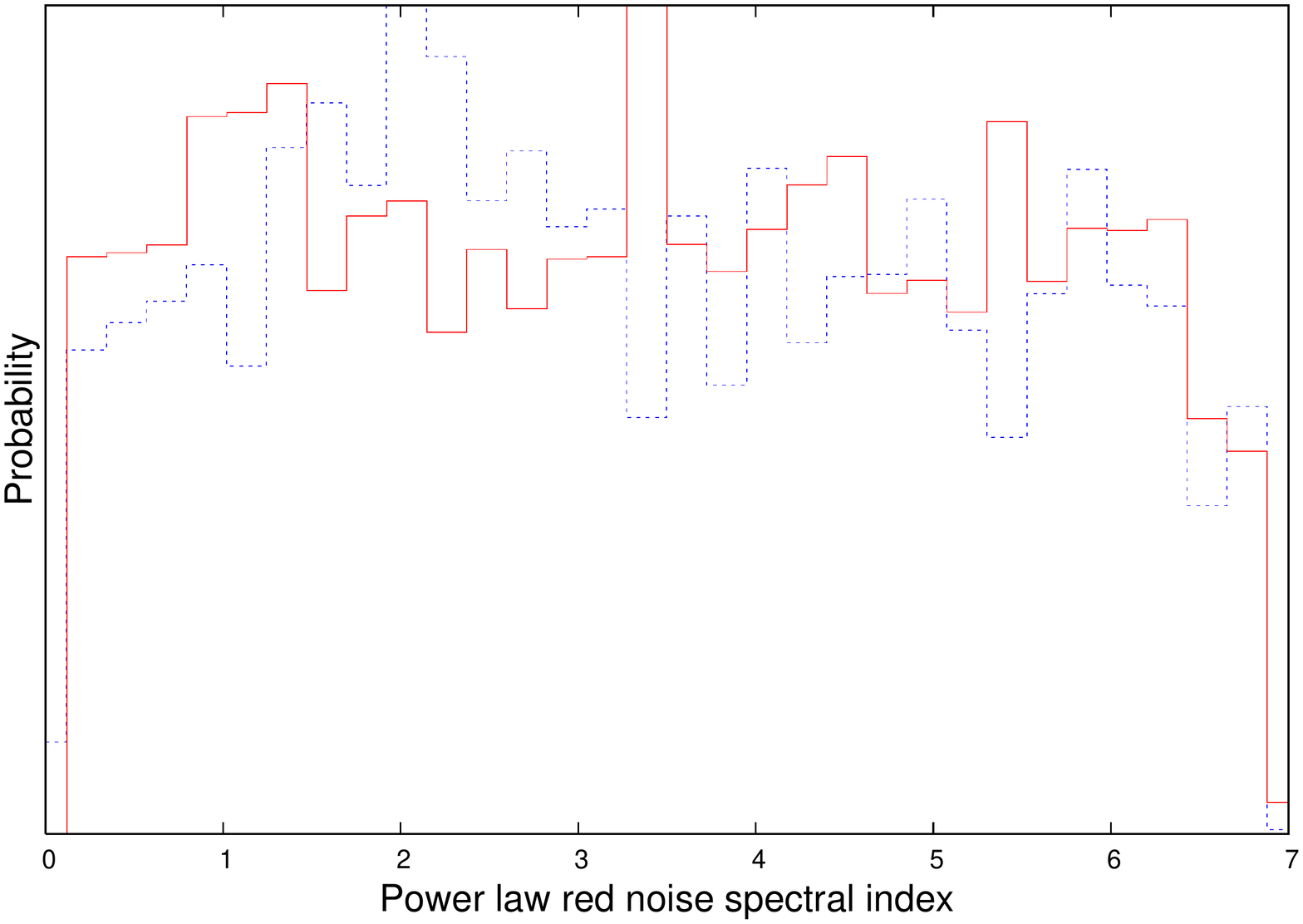} \\
\end{array}$
\end{center}
\vspace{-1cm}
\caption{1-dimensional marginalised posteriors for the red spin noise amplitude (left) and spectral index (right) for a log uniform prior on the amplitude when including (blue),  and not including (red),  the simulated DM observations as additional prior information. In both cases the red noise signal is totally unconstrained below some upper limit, however when using a log uniform prior that upper limit is dependent on the lower limit of the prior.  We therefore repeat the analysis using a uniform prior on the red noise amplitude in order to obtain a robust upper limit on the signal.\label{Fig:LogRed}}
\end{figure*}

In addition to the timing model and DM parameters, we also include a red noise power law model in our analysis, of the same functional form as the DM spectrum.  We initially performed our analysis using a log uniform prior on the red noise amplitude, however as can be seen in Fig. \ref{Fig:LogRed} the signal is completely unconstrained below some upper limit.  When using a log uniform prior this upper limit will be dependent upon the lower limit chosen for the prior, and as such we instead use a uniform prior on the red noise amplitude, in order to obtain a robust upper limit on the signal.

\begin{figure*}
\begin{center}$
\begin{array}{ccc}
%
\includegraphics[width=50mm]{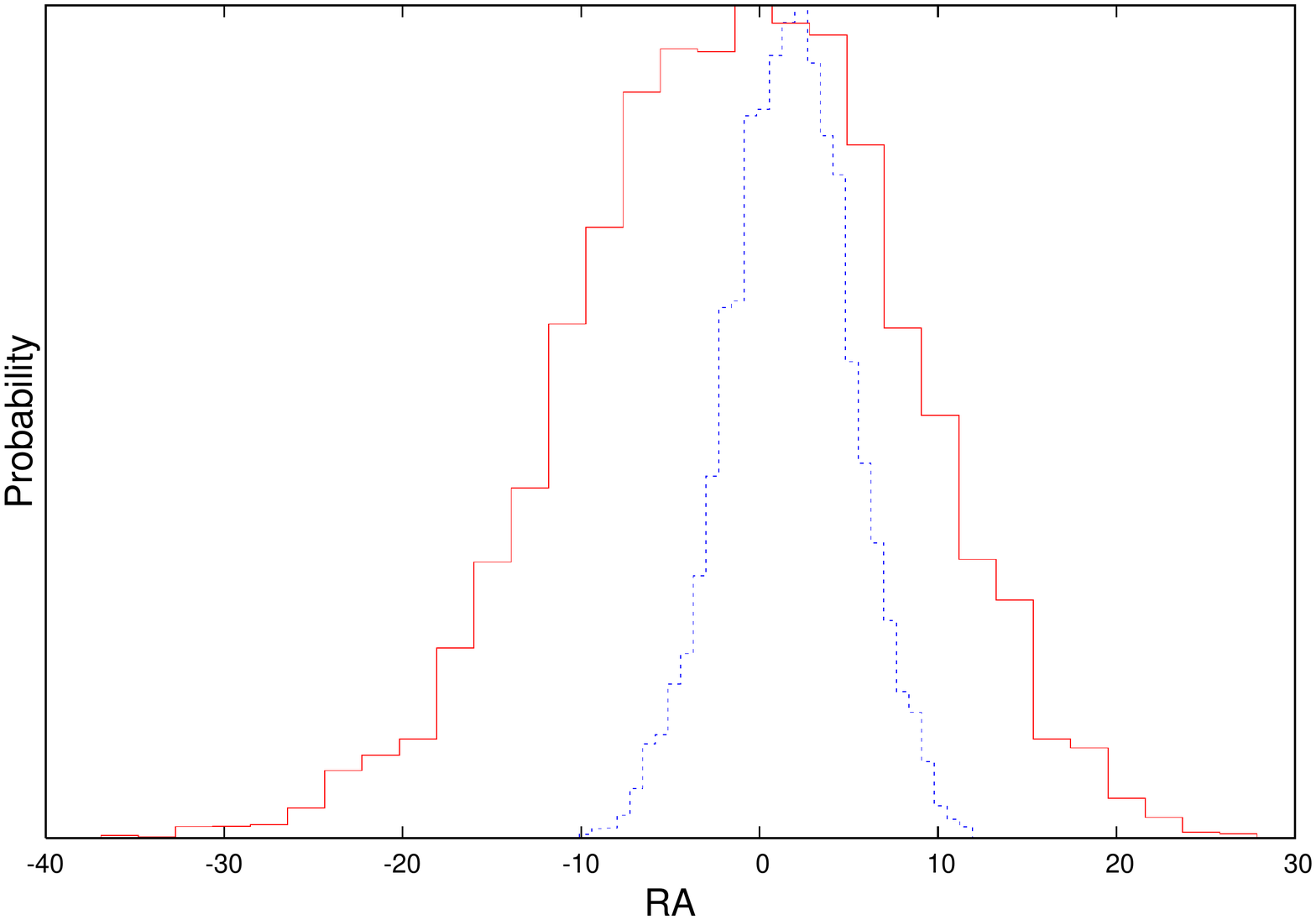} &
\includegraphics[width=50mm]{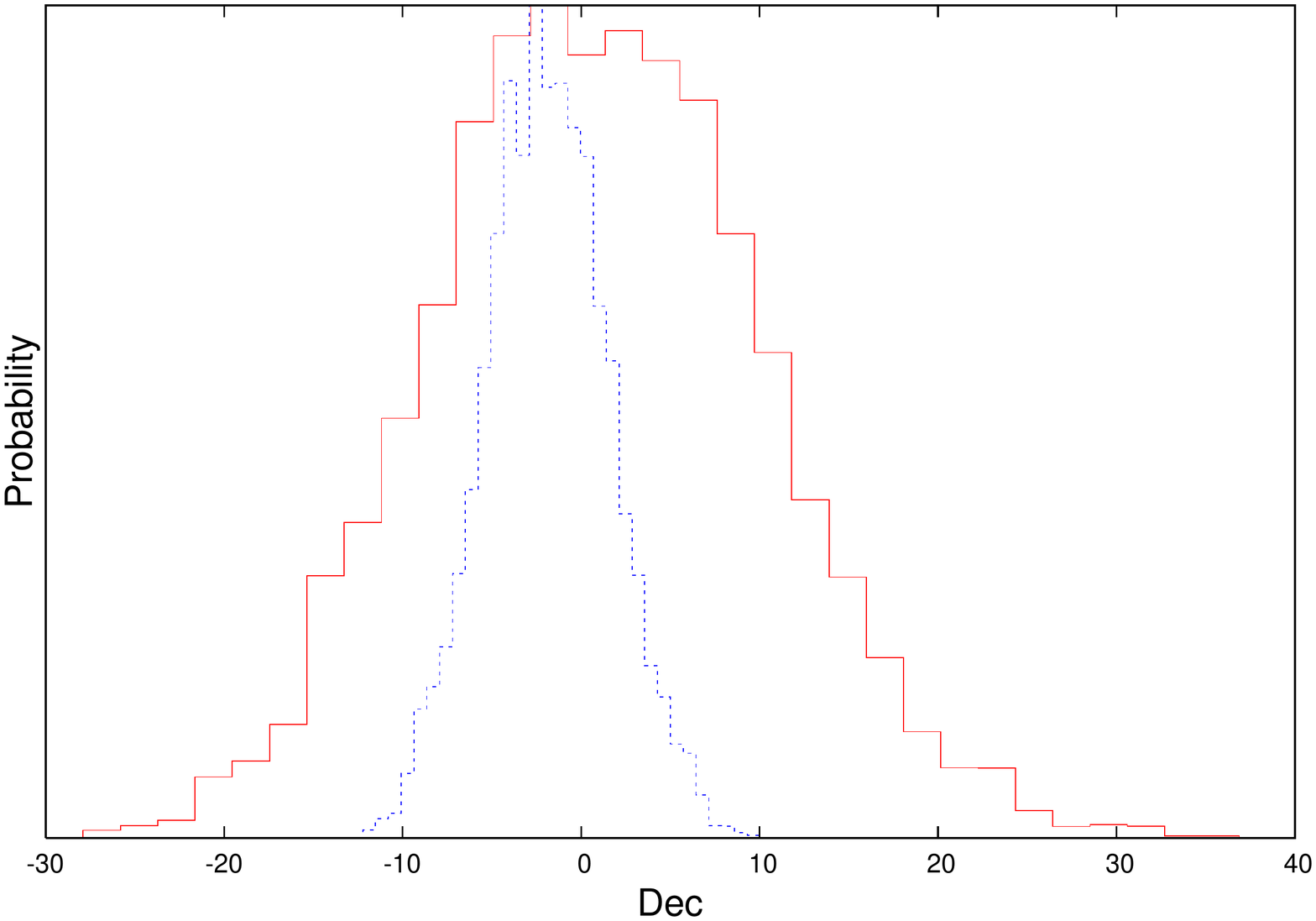} &
\includegraphics[width=50mm]{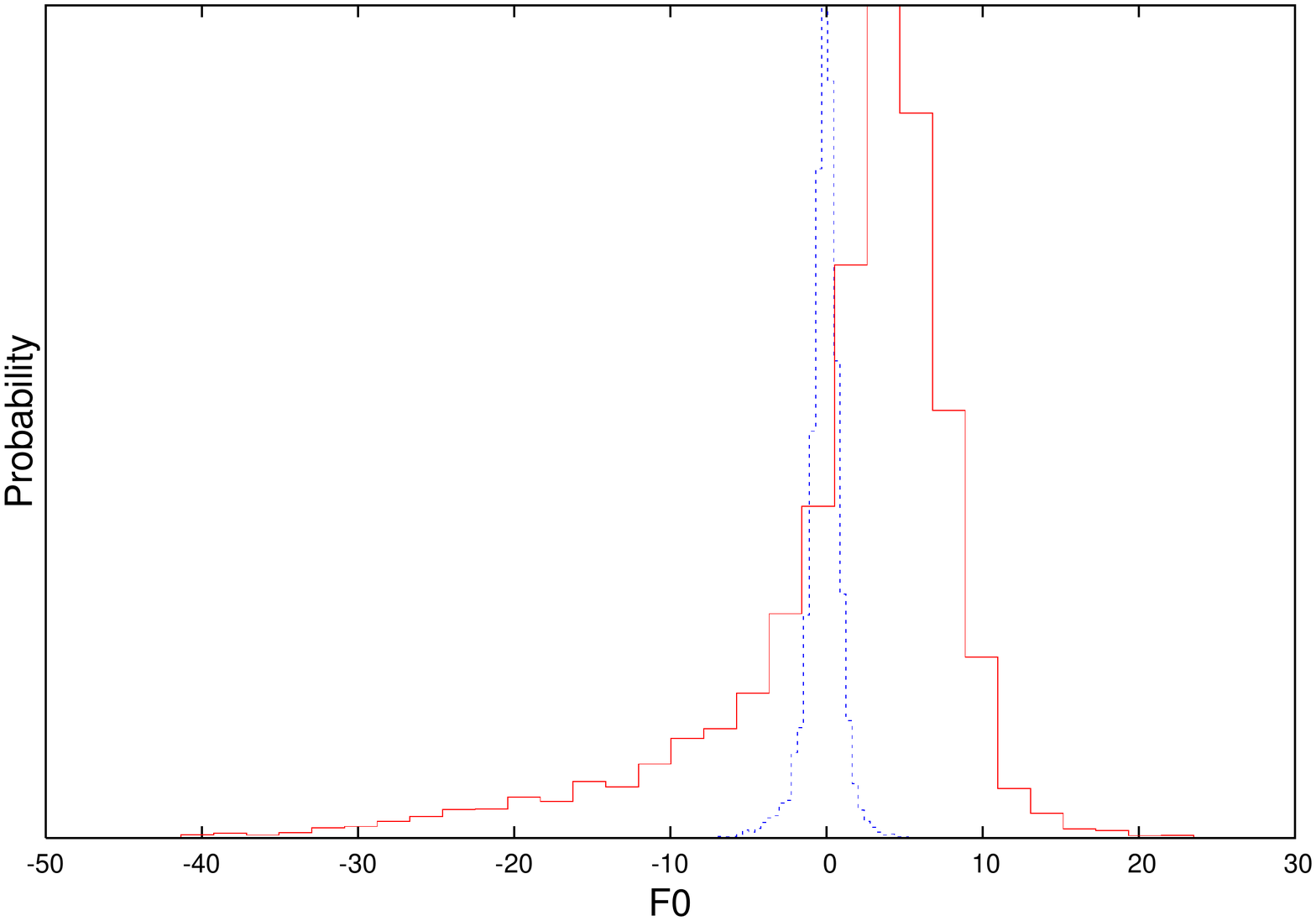} \\
%
\includegraphics[width=50mm]{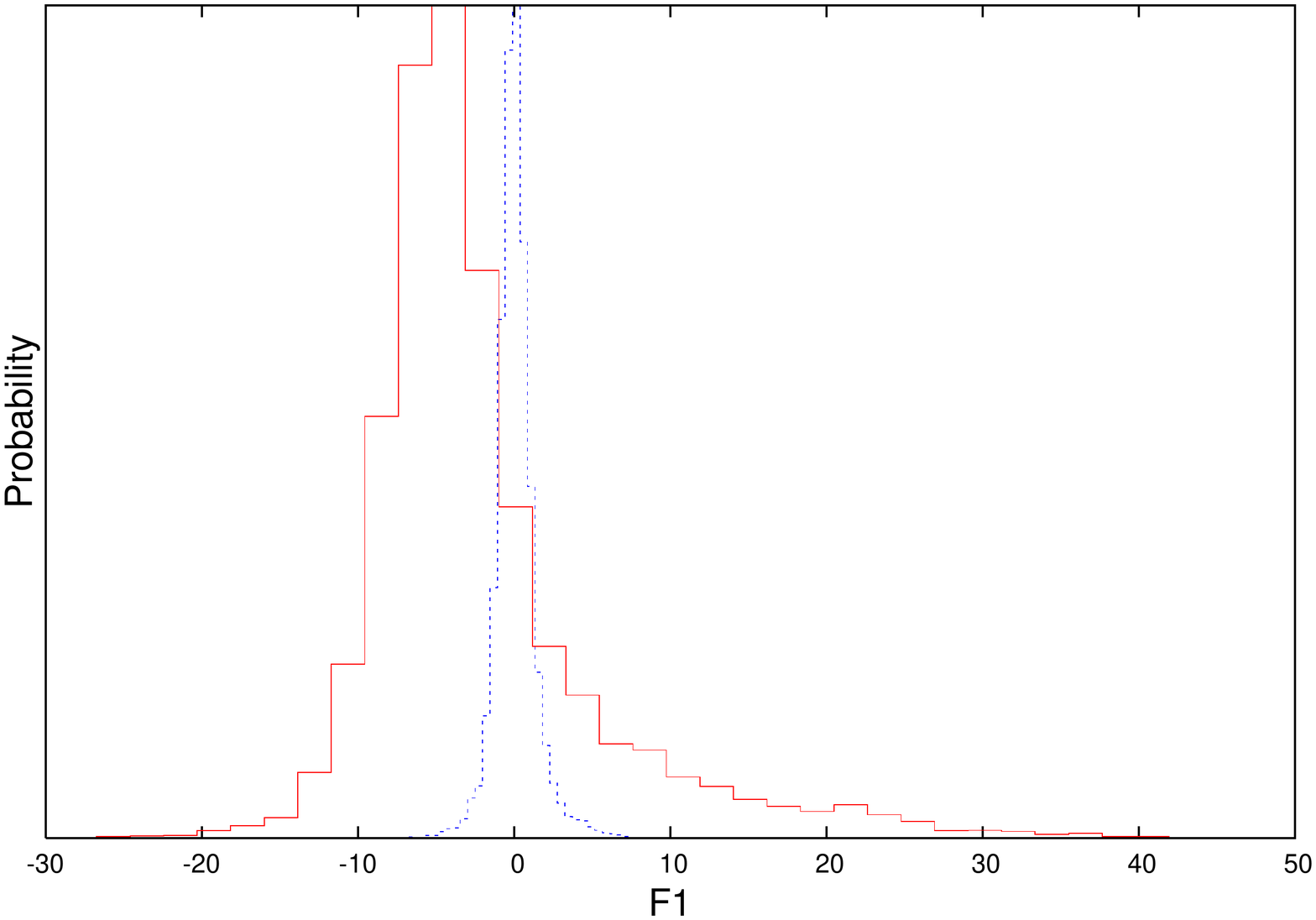} &
\includegraphics[width=50mm]{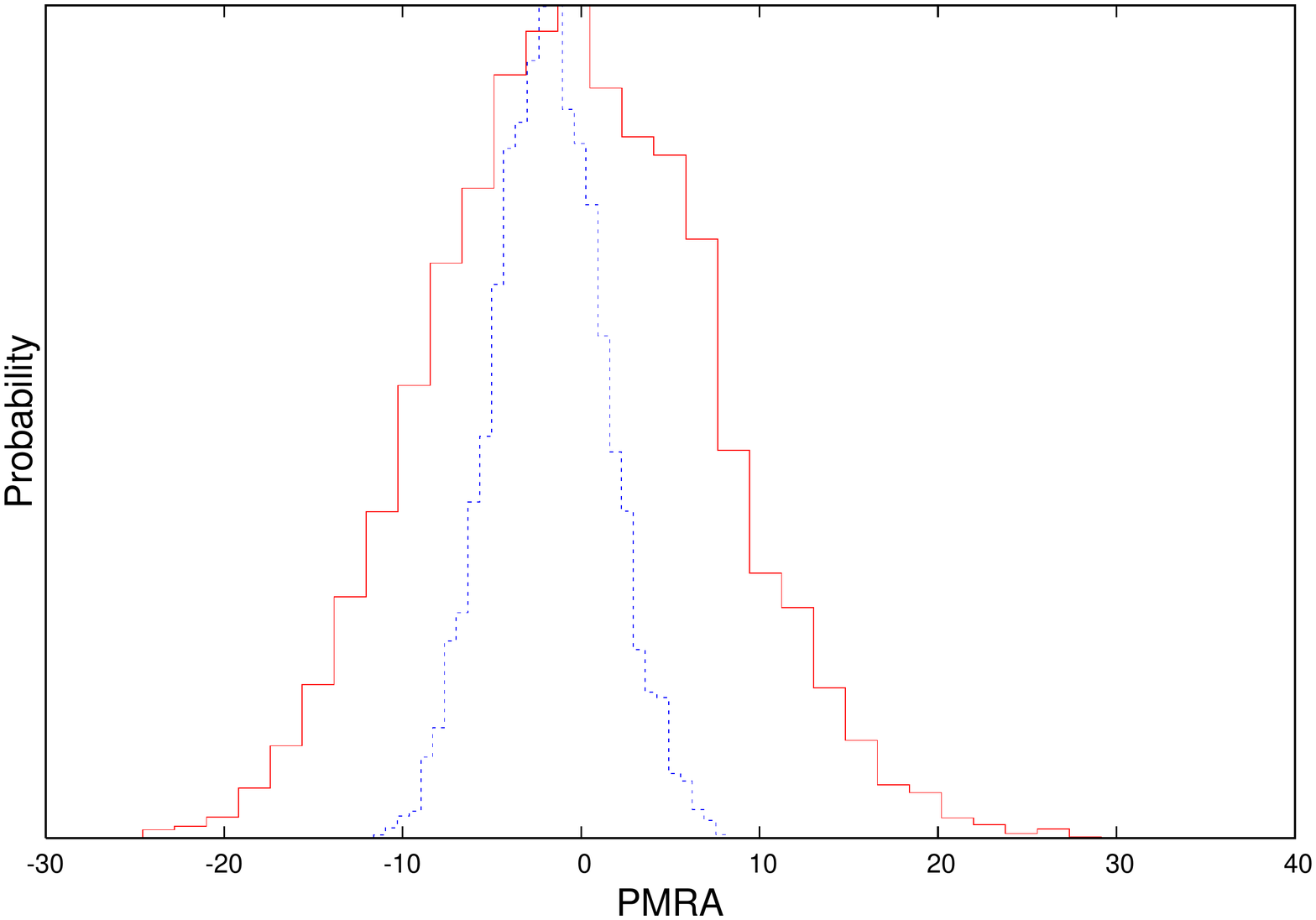} &
\includegraphics[width=50mm]{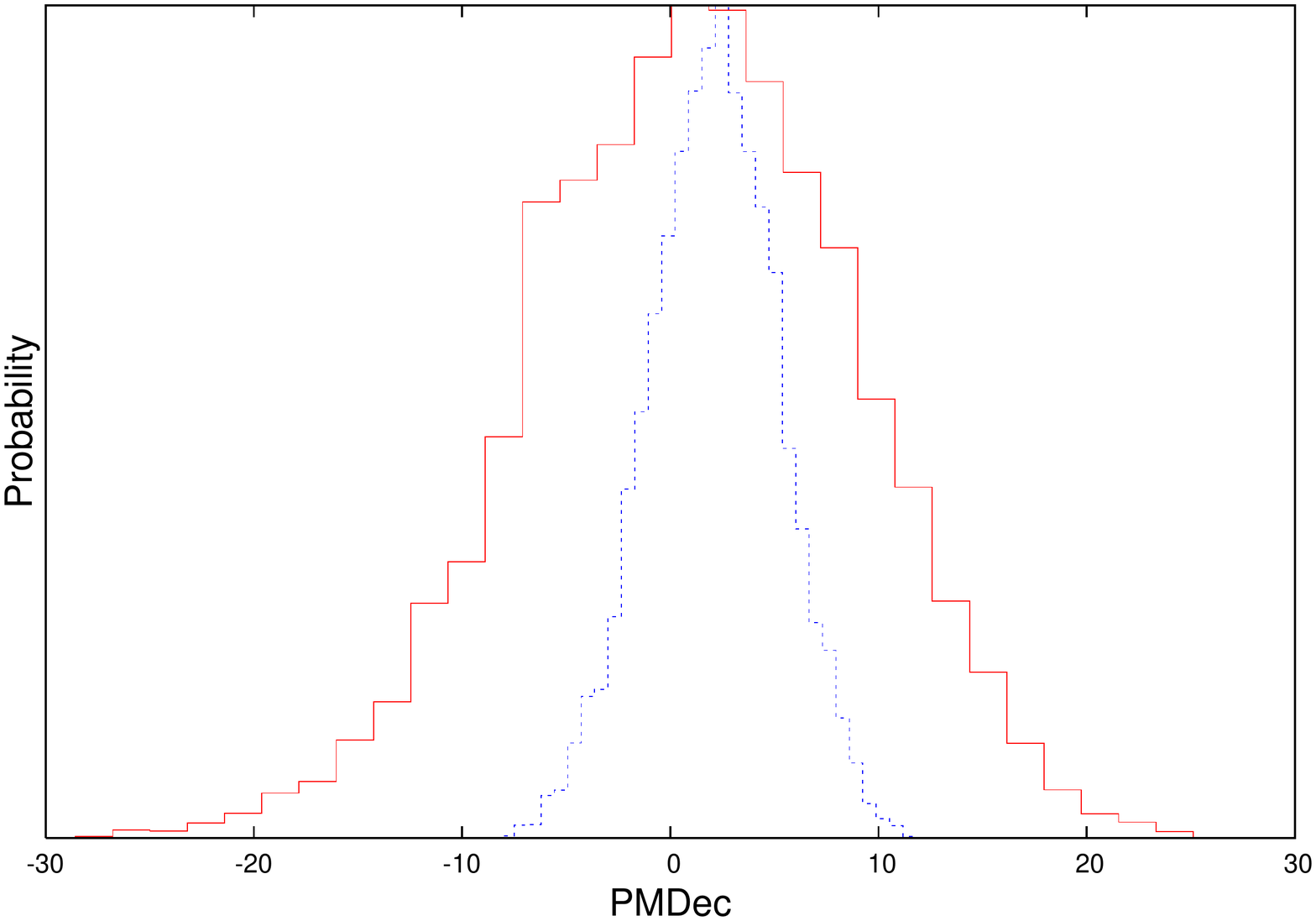} \\
%
\includegraphics[width=50mm]{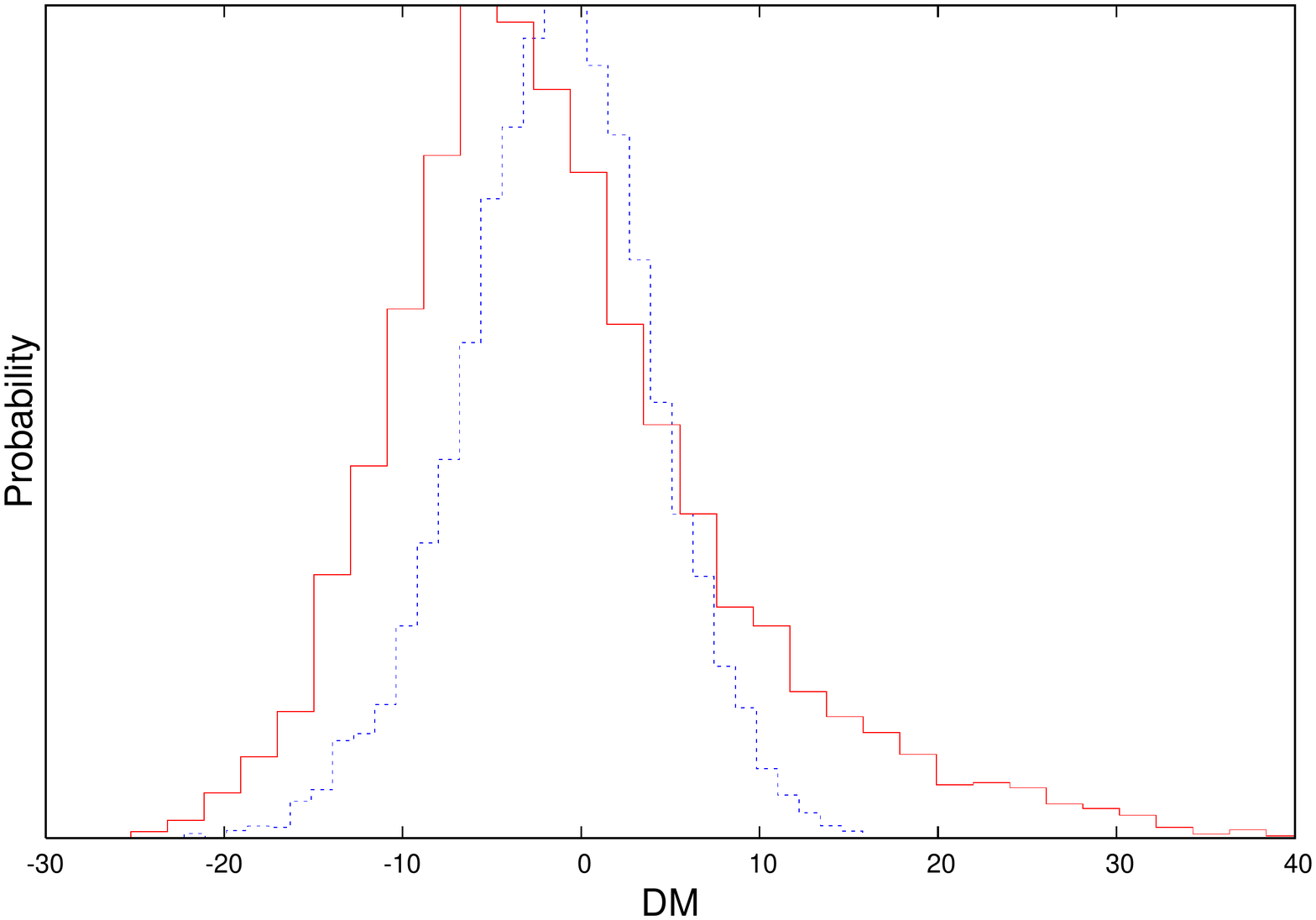} &
\includegraphics[width=50mm]{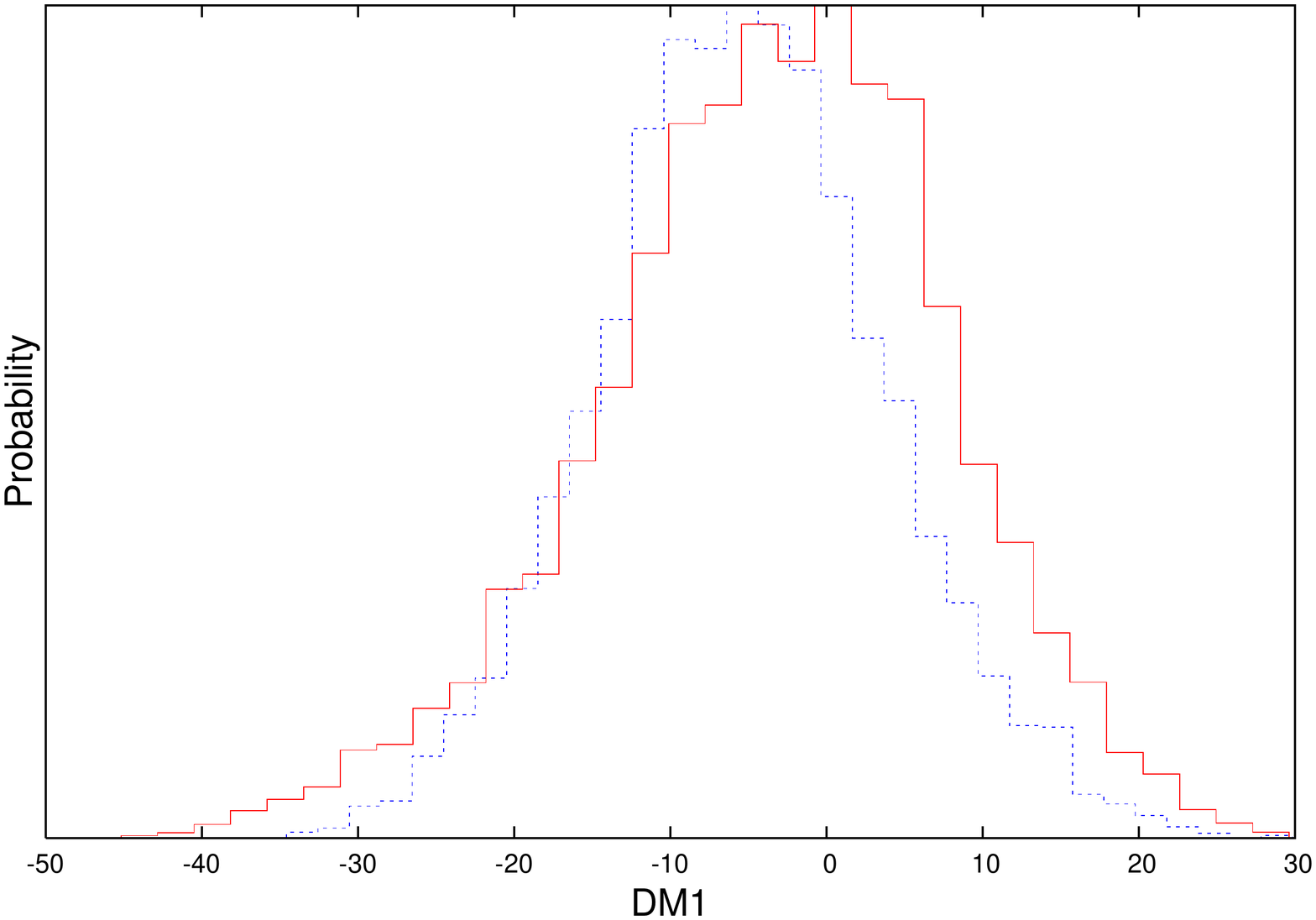} &
\includegraphics[width=50mm]{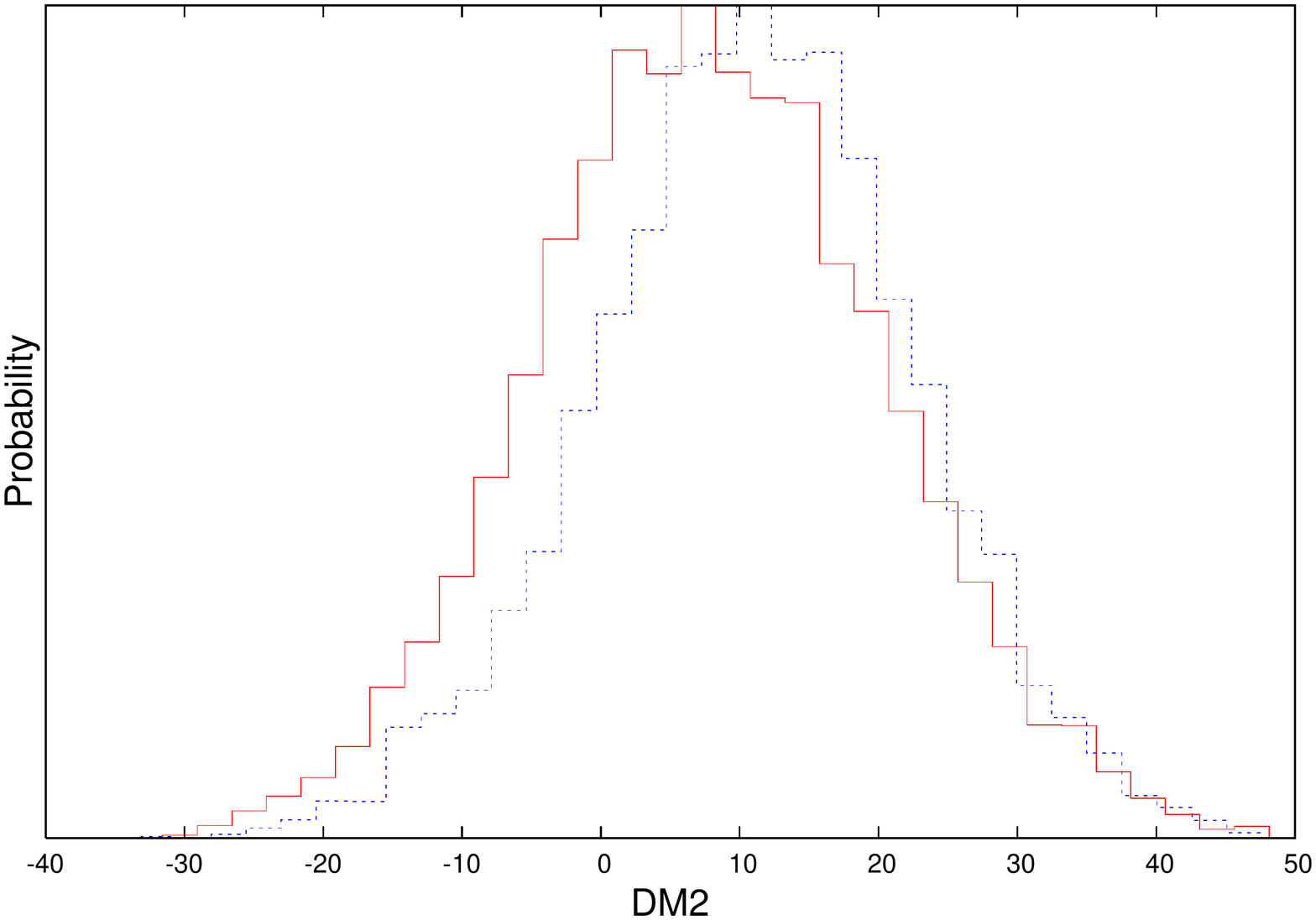} \\
%
\includegraphics[width=50mm]{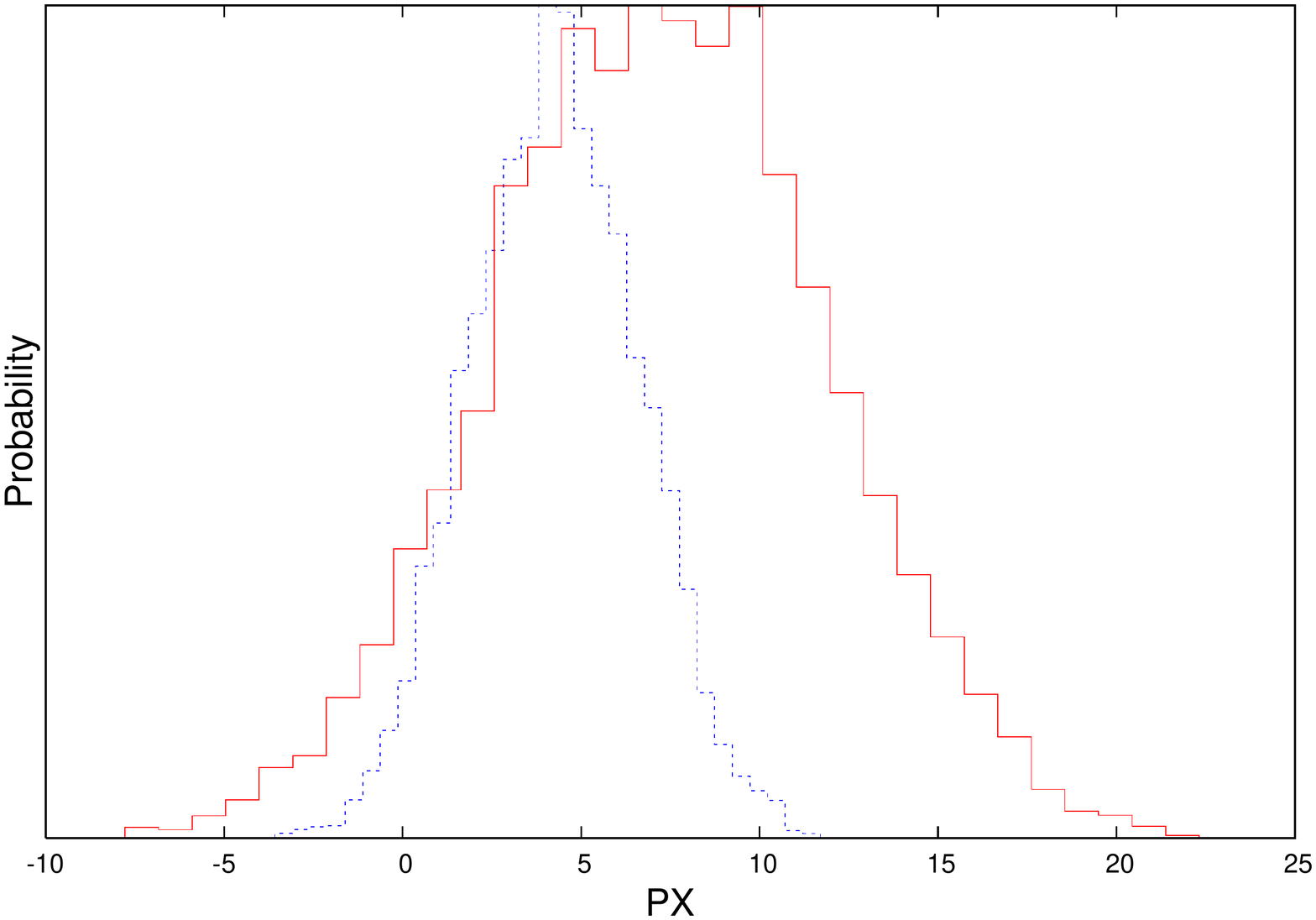} &
\includegraphics[width=50mm]{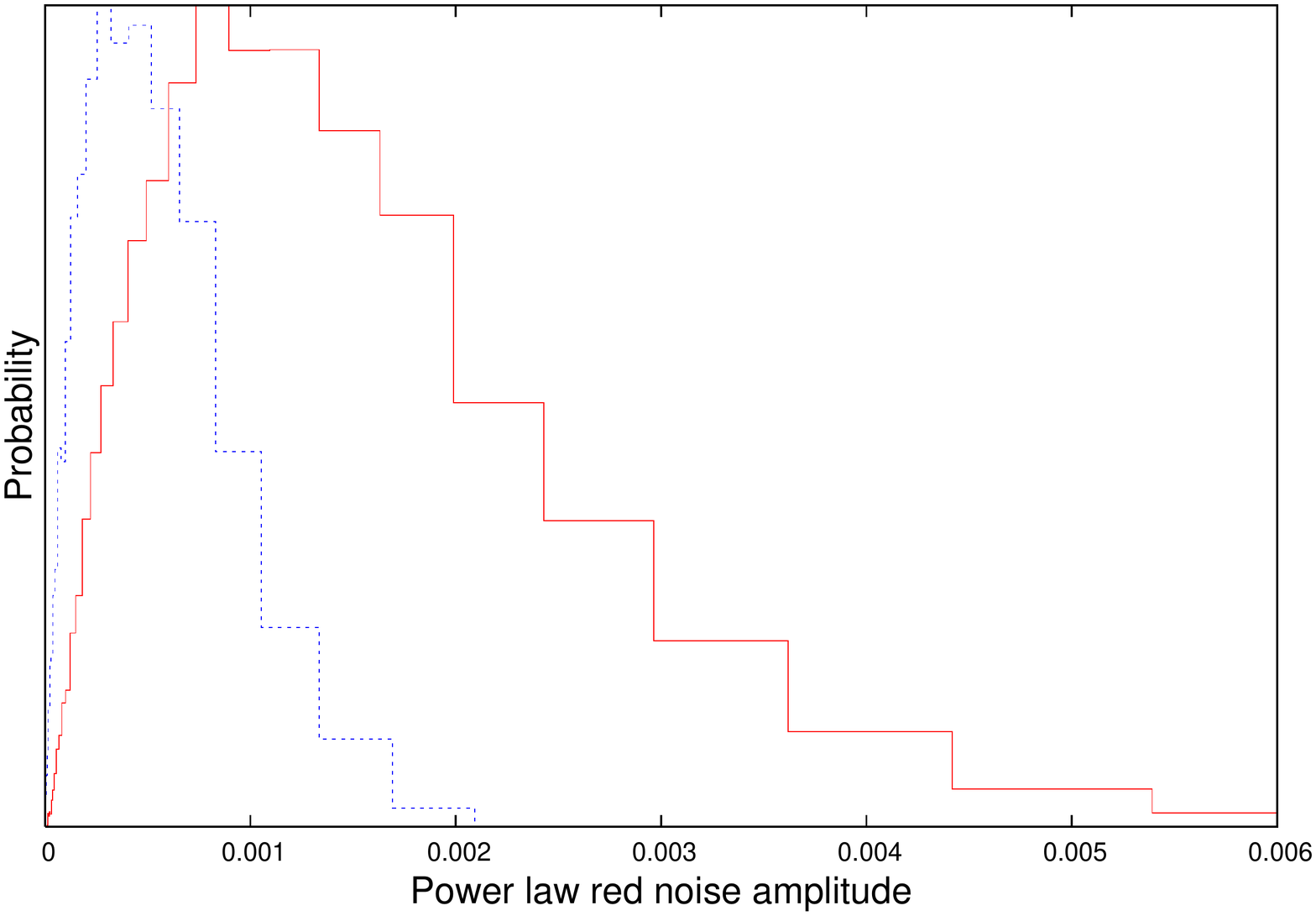} &
\includegraphics[width=50mm]{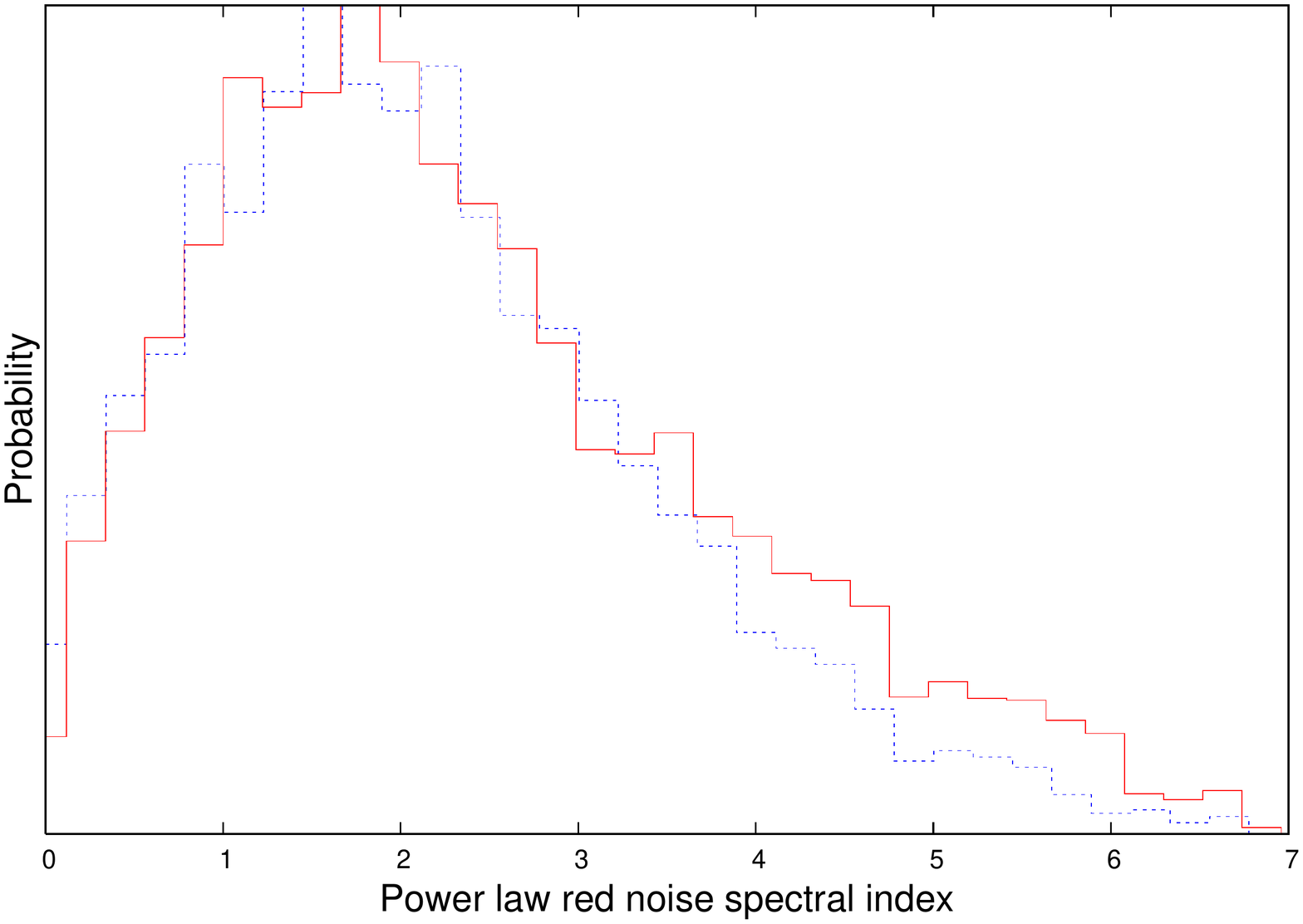} \\
%
\includegraphics[width=50mm]{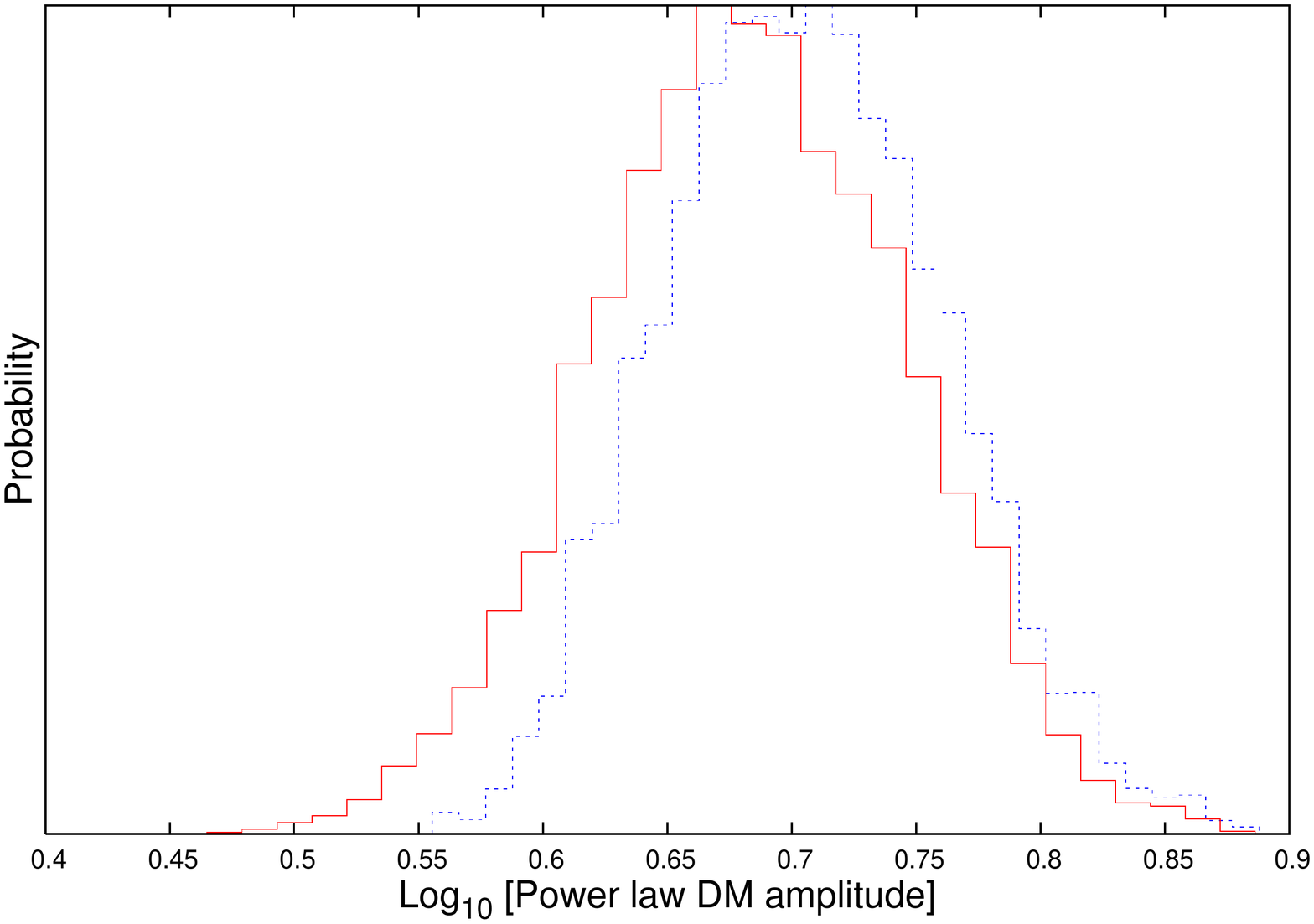} &
\includegraphics[width=50mm]{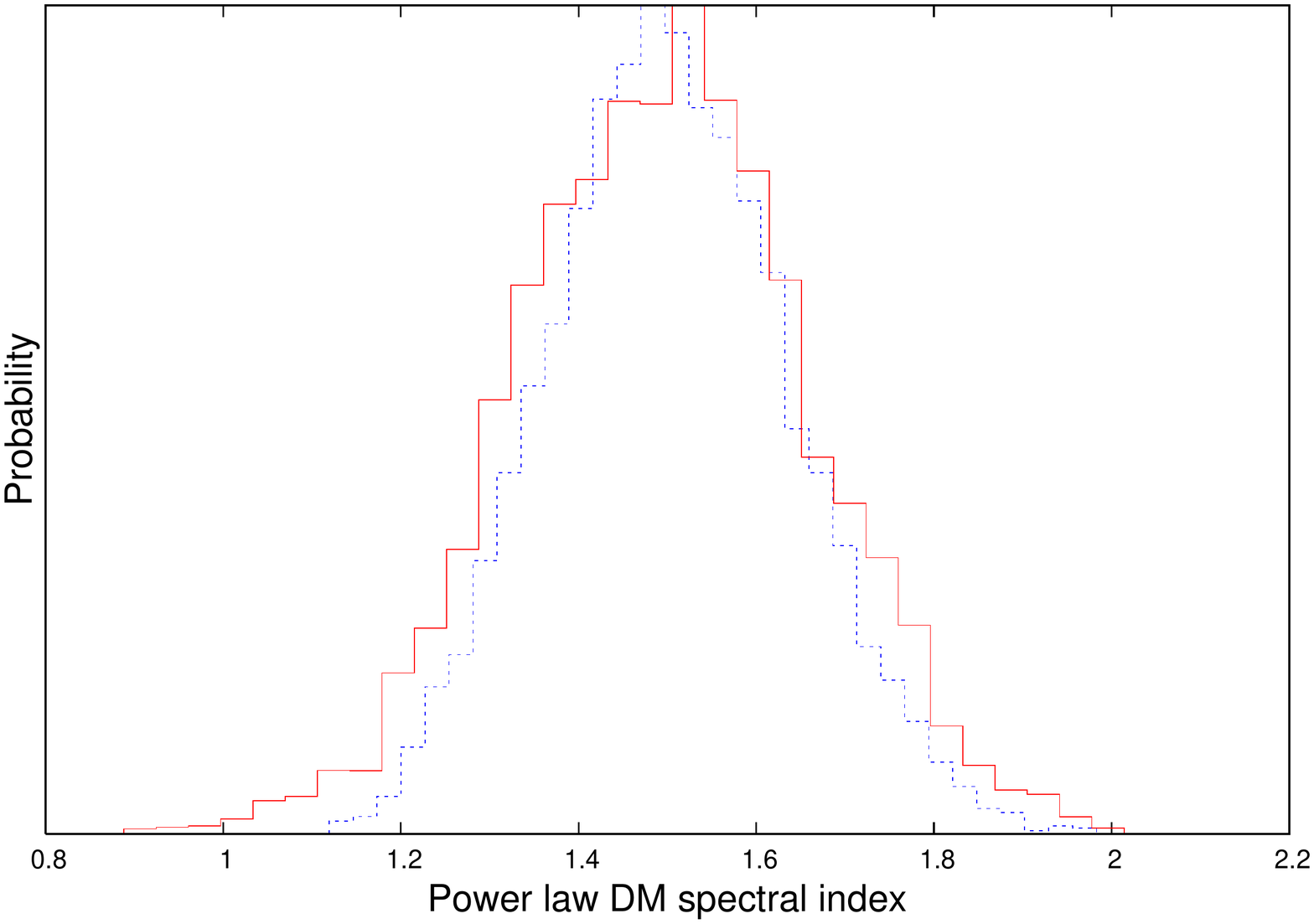} & \\
\end{array}$
\end{center}
\vspace{-0.8cm}
\caption{1-dimensional marginalised posteriors for the timing model and stochastic parameters in simulated data of PSR J0030+0451 when including (blue),  and not including (red),  the simulated DM observations as additional prior information. For those timing model parameters listed in Table \ref{Table:Sim1} the injected value is at 0 on the $x$ axis, which is given in units of the 1$\sigma$ uncertainty returned by Tempo2 when performing the fit when not including either the red noise or DM power law model components. All timing model parameters show substantial increase in the precision with which they are recovered when including the additional prior information, from a factor $\sim$ 2, up to a factor $\sim 7$ in the case of F0.  The 2$\sigma$ upper limit on the red noise amplitude also decreases by a factor $\sim$ 3 when including this prior information from $10^{-2.48}$ to $10^{-2.95}$ \label{figure:TimingPosteriors}}
\end{figure*}

In Fig. \ref{figure:TimingPosteriors} we show the one dimensional posteriors for the timing model and stochastic parameters given in Table \ref{Table:Sim1} when including (blue),  and not including (red),  the simulated DM observations as additional prior information. For the timing model parameters the injected value, when known,  is given by 0 on the $x$ axis, which is in units of the 1$\sigma$ uncertainty returned by Tempo2 when not including either the red noise or DM power law model components. The clear result here is that the precision with which the timing model parameters have been recovered has improved significantly when including the additional prior information, between a factor $\sim 2-7$.

Comparing the posteriors for the amplitude of the red noise power law, when including the additional prior information the upper limit decreases by a factor of $\sim 3$, demonstrating how critical such data will be in constraining gravitational wave signals in pulsar timing data.

\section{Conclusions}
\label{Section:Conclusion}

We have discussed a method of including discrete measurements of the dispersion measure in the direction of a pulsar as prior information in the analysis of that pulsar.   By using an existing Bayesian framework, this prior information can be simply folded into the analysis and used to constrain the DM signal realisation in pulsar timing data.  We have shown that this method can be applied where no multi-frequency data exists across much of the dataset, and does not require simultaneous multi-frequency data to be present for any observing epoch.

We have shown that, as expected, including this prior information can greatly increase both the precision of the timing model parameters recovered from the analysis, as well as increase the sensitivity to red noise in the data.  Clearly the level of improvement in real data will be entirely dependent on the dataset in question, however the inclusion of such prior information will likely prove extremely useful both to test the validity of existing models for DM, and to improve constraints in future analysis.

\section{Acknowledgements}

Many thanks to Jason Hessels for assistance in the production of this work.

\label{lastpage}

\end{document}